\documentclass[a4paper]{jpconf}

\usepackage{graphicx}
\usepackage{color}
\usepackage{psfrag}
\usepackage{amsmath,amssymb}

\newcommand{\bea}{\begin{eqnarray}}
\newcommand{\beas}{\begin{eqnarray*}}

\newcommand{\eea}{\end{eqnarray}}
\newcommand{\eeas}{\end{eqnarray*}}
\newcommand{\bd}{\begin{displaymath}}
\newcommand{\ed}{\end{displaymath}}
\newcommand{\be}{\begin{equation}}
\newcommand{\ee}{\end{equation}}

\newcommand{\RE}{{\rm Re}}
\newcommand{\IM}{{\rm Im}}

\newcommand{\gev}{\mbox{GeV}}

\newcommand{\Br}{{\rm Br}}

\newcommand{\rC}{r_{\textrm{C}}}

\newcommand{\rEW}{r_{\textrm{EW}}}

\newcommand{\rtEW}{\tilde{r}_{\textrm{EW}}}

\newcommand{\rtEWA}{\tilde{r}_{\textrm{EW}}^{\textrm{A}}}
\newcommand{\rtEWi}{\tilde{r}_{\textrm{EW},\,i}}
\newcommand{\rtEWCi}{\tilde{r}_{\textrm{EW},\,i}^{\textrm{C}}}
\newcommand{\rtEWAi}{\tilde{r}_{\textrm{EW},\,i}^{\textrm{A}}}

\newcommand{\captionfonts}{\small}
\makeatletter  
\long\def\@makecaption#1#2{%
  \vskip\abovecaptionskip
  \sbox\@tempboxa{{\captionfonts #1: #2}}%
  \ifdim \wd\@tempboxa >\hsize
    {\captionfonts #1: #2\par}
  \else
    \hbox to\hsize{\hfil\box\@tempboxa\hfil}%
  \fi
  \vskip\belowcaptionskip}
\makeatother   

\begin{document}

\begin{flushright}
TTP11-10\\
MZ-TH/11-10\\
February 2011
\end{flushright}

\vskip-1.5cm

\title{Probing new physics in electroweak penguins through $B_d$ and $B_s$ decays}

\author{Lars Hofer$^{1,2}$, Dominik Scherer$^1$,
and Leonardo Vernazza$^{3,}$\footnote[4]{Alexander-von-Humboldt fellow.}}

\address{$^1$ Institut f\"ur Theoretische Teilchenphysik,
Karlsruhe Institute of Technology, D--76128 Karlsruhe, Germany}
\address{$^2$ Institut f\"ur Theoretische Physik und Astrophysik,
Universit\"at W\"urzburg, D--97074 W\"urzburg, Germany}
\address{$^3$ Institut f\"ur Physik (THEP), Johannes Gutenberg-Universit\"at, D--55099 Mainz, Germany}

\ead{lars.hofer@physik.uni-wuerzburg.de,\,dominik@particle.uni-karlsruhe.de,\\
     vernazza@uni-mainz.de}

\begin{abstract}
An enhanced electroweak penguin amplitude due to the presence of unknown new
physics can explain the discrepancies found between theory and experiment
in the $B\to\pi K$ decays, in particular in
\mbox{$A_{\rm CP}(B^-\to\pi^0 K^-)-A_{\rm CP}(\bar{B}^0\to\pi^+ K^-)$},
but the current precision of the theoretical and experimental results does
not allow to draw a firm conclusion. We argue that the
$\bar B_s\to\phi\rho^0$ and $\bar B_s\to\phi\pi^0$ decays offer an additional tool
to investigate this possibility. These purely isospin-violating decays are
dominated by electroweak penguins and we show that in presence of a new
physics contribution their branching ratio can be enhanced by about
an order of magnitude, without violating any constraints from
other hadronic $B$ decays. This makes them very interesting modes for LHCb and future
$B$ factories. In \cite{Hofer:2010ee} we have performed both a model-independent analysis and a
study within realistic New Physics models such as a modified-$Z^0$-penguin scenario,
a model with an additional $Z^{\prime}$ boson and the MSSM. In this article we summarise
the most important results of our study.
\end{abstract}

\section{Introduction}

The four $B\to\pi K$ decays have been very useful in testing flavour structure
and CP violation in the Standard Model (SM) since the late 1990s. These decays
have a small branching ratio, $\mathcal O(10^{-6})$, and are sensitive
to New Physics (NP) contributions. In the past years some discrepancies between
$B\to\pi K$ measurements and SM predictions have occurred, provoking
speculations on a ``$B\to\pi K$ puzzle''. To date, the measurements of the
branching fractions have fluctuated towards the SM predictions, the latter still
suffering from large hadronic uncertainties, and only the following difference
of CP asymmetries shows an unexpected behavior
\cite{DeltaACP}:
\be\label{a1}
\Delta A_{\rm CP} \equiv A_{\rm CP}(B^-\to\pi^0 K^-)
-A_{\rm CP}(\bar{B}^0\to\pi^+ K^-)
\stackrel{\text{SM}}{=} 1.9^{+5.8}_{-4.8}\,\%
\stackrel{\text{exp.}}{=} (14.8 \pm 2.8)\, \%.
\ee
The theory result, derived within the framework of QCD factorisation (QCDF)
\cite{QCDF}, and the experimental result, taken from
\cite{Barberio:2008fa}, differ by $2.5\sigma$. This result suggests a
violation of the isospin symmetry beyond the amount expected in the SM.
It can be interpreted as a hint of a NP electroweak (EW) penguin amplitude,
or alternatively as an enhanced low-energy hadronic effect
which the theory is not able to catch. Present data do not allow to
distinguish between these two possibilities.

Our aim is to show that the question can be partially addressed exploiting the
large variety of non-leptonic $B$ decays into two charmless mesons. For this
purpose, in \cite{Hofer:2010ee,Hofer:2009ct} we have studied the \textit{purely}
isospin-violating decays $\bar B_s\rightarrow \phi \rho^0$ and
$\bar B_s\rightarrow \phi \pi^0$, which are dominated by EW penguins, as
pointed out for the first time in \cite{Fleischer}. We have shown that
if NP in this sector exists at a level where it can explain the
$\Delta A_{\rm CP}$ puzzle, it could be clearly visible in these purely
isospin-violating decays. We present here a summary of the results,
referring to \cite{Hofer:2010ee} for a more detailed discussion.

\section{\boldmath
Isospin violation in $B\to\pi K$ and
$B_s\to\phi \rho^0$, $B_s\to \phi \pi^0$ decays}

The $B\to \pi K$ decays are dominated by the isospin-conserving QCD penguin
amplitude. Nevertheless, they receive small contributions from the tree and the
EW penguin amplitude, which are isospin-violating. Combining
measurements of the four different decay modes $B^-\to \pi^-\bar{K}^0$, $B^-\to
\pi^0K^-$, $\bar{B}^0\to \pi^+K^-$ and $\bar{B}^0\to \pi^0\bar{K}^0$, it is
possible to construct observables in which the leading contribution from the QCD
penguin drops out, so that they are sensitive to isospin violation and therefore
to a possible new contribution with the structure of the EW penguins.

The decay amplitudes can be written in terms topological amplitudes $r_i$ which are normalised to
the dominant QCD penguin contribution. Introducing as well a NP
EW penguin amplitude, such that
\be\label{g8}
r_\text{EW}  \rightarrow r_\text{EW} + \tilde{r}_\text{EW}
e^{-i\delta},\hspace{1.0cm}
r_\text{EW}^{\textrm{C}}  \rightarrow r_\text{EW}^{\rm C} +
\tilde{r}_\text{EW}^{\rm C} e^{-i\delta}, \hspace{1.0cm}
r_\text{EW}^{\textrm{A}}  \rightarrow r_\text{EW}^{\rm A} +
\tilde{r}_\text{EW}^{\rm A} e^{-i\delta},
\ee
one obtains
\be\label{g10}
\Delta A_{\rm CP}\,\simeq\,-2\,\IM \left(r_{\rm C}\right) \sin\gamma +
2\,\IM\left(\tilde{r}_\text{EW} +
\tilde{r}_\text{EW}^\textrm{A}\right)\sin\delta\,.
\ee
Here $r_\text{EW}$, $r_\text{EW}^{(\textrm{C},\textrm{A})}$
represent respectively the SM colour-allowed, colour-suppressed and
annihilation electroweak penguin amplitudes, while the
$\tilde{r}_\text{EW}^i$ are the corresponding NP contributions. It is
clear from (\ref{g10}) that a large $\Delta A_{\rm CP}$ can be obtained
both through enhanced low-energy effects in $\IM \left(r_{\rm C}\right)$,
as well as by means of a NP contribution. An $\IM \left(r_{\rm C}\right)$ large
enough to solve the $\Delta A_{\rm CP}$ discrepancy, however, would
break QCDF, so that within this framework a NP solution is favoured.
Although one can construct many other observables which are sensitive to
isospin violation, the colour-allowed EW
penguin contribution and the colour-suppressed tree contribution
cannot be disentangled within the $B\to \pi K$ decays alone
since they enter the amplitudes exclusively in the combination
\begin{equation}\label{eq:ParaCombi}
    \rEW\,-\,\rC\,e^{-i\gamma}.
\end{equation}

Our proposal is now to get a more complete picture by
considering as well the decays
\begin{center}
\fbox{ $\bar{B}_s \to \phi \rho^0 \qquad \text{and} \qquad \bar B_s \to \phi
\pi^0$.}
\end{center}
These decays are purely isospin-violating and their structure in the SM
is very simple. Even though we
are facing the same type of EW penguin vs colour-suppressed tree amplitude
pollution as in the $B\to\pi K$ decays, various elements suggest that
the analysis of these decays may be interesting:
\begin{itemize}
 \item these decays are dominated by the EW penguin amplitude
 \cite{Fleischer}, such that a NP effect coming from a new EW penguin amplitude should
 have more spectacular effects in these decays than in the QCD penguin dominated $B\to\pi K$:
 A new EW amplitude of the same order as the SM one can enhance the branching ratios by about an
order of magnitude.
 \item the nature of these $B_s$ decays is different from the
$B\to \pi K$ decays, the first being $PV$, $VV$ decays, respectively,
while the latter are $PP$ decays. The non-perturbative low-energy QCD
dynamics is expected not to be correlated among the two classes of decays,
and the exact relation cannot be determined within QCDF. Phenomenological
comparison with other decays such as $B\to \pi\pi$, $B\to\pi\rho$,
$B\to\rho\rho$ suggests a decreasing of non-perturbative effects when
vector mesons appear in the final state. A NP contribution is
instead of high-energy origin, and its effects can be reliably studied
within perturbation theory, allowing for a correlation among the
$B\to \pi K$ and the $B_s$ decays of interest.
\end{itemize}

\section{Model-independent analysis}

\begin{figure}[t]
\begin{center}
  \includegraphics[width=0.75\textwidth]{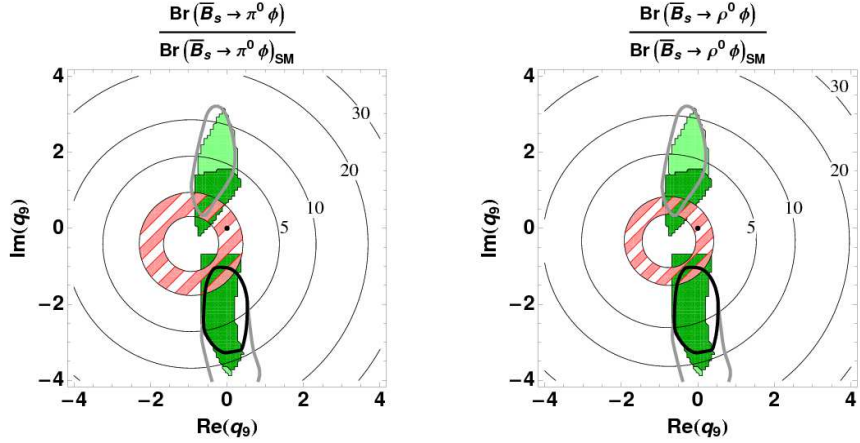}
\end{center}
  \caption{Enhancement factors of the $\bar{B_s}\rightarrow~\phi\rho^0,\phi\pi^0$
  branching ratios with respect to their SM values. The black dot represents the
  SM result while the red striped region shows the theoretical uncertainty
  in the SM. The dark green area is the region allowed by the $2\,\sigma$
  constraints from $\bar B \to \pi K^{(*)},\rho K^{(*)},\phi K^{(*)}$
  and $\bar B_s\to\phi\phi,\bar K K$ decays; for comparison, the light green area represents
  the area allowed by constraints from isospin-sensitive observables only, considering
  only $\bar B \to \pi K,\pi K^{(*)},\rho K$ decays. The solid black line
  represents the $1\sigma$ CL of the fit with $S_{CP}(\bar B^0 \to \pi \bar{K}^0)$,
  while the solid grey line represents the $1\sigma$ CL of the fit without it.
  Here the scenario $q_9\neq 0$ is shown.}
  \label{figq7q9}
\end{figure}

\begin{figure}[t]
\begin{center}
  \includegraphics[width=0.75\textwidth]{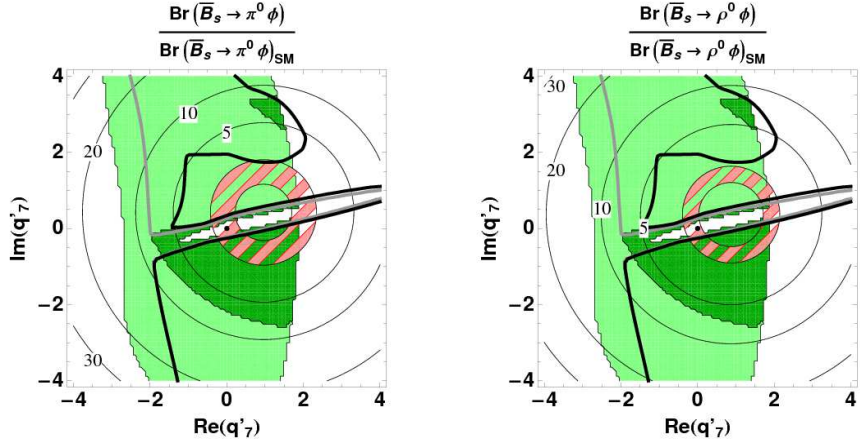}
\end{center}
  \caption{Enhancement factors of the $\bar{B_s}\rightarrow~\phi\rho^0,\phi\pi^0$
  branching ratios with respect to their SM values. The meaning of the contours and regions
  is the same as in fig.~\ref{figq7q9}. Here the scenario $q_7^{\prime}\neq0$
  Is displayed. The $1\sigma$ region of the fit is the
  region to the right of the black (grey) curve.}
  \label{figq7q9P}
\end{figure}

We first perform a model-independent analysis. We parameterise NP in EW
penguins by adding corresponding terms $C_{7,9}^{(\prime)\rm NP}$ to the Wilson coefficients
of the effective weak Hamiltonian,
\be\label{g23}
C_{7,9}^{(\prime)\rm NP}(m_W) = C_9^{\rm LO}(m_W)\,q_{7,9}^{(\prime)},
\qquad \qquad q_{7,9}^{(\prime)}  = |q_{7,9}^{(\prime)}|e^{i
\phi_{7,9}^{(\prime)}}.
\ee
Here $C_9^{\rm LO}$ denotes the leading-order SM coefficient, the primed operators are
obtained from the SM $Q_{7,9}$ by flipping the chiralities of the quark fields.
The purpose is to perform a $\chi^2$-fit in order to determine
the NP parameters in such a way that they describe better $B\to \pi K$ data. In particular
we look for a solution of the $\Delta A_{\textrm{CP}}$ discrepancy. Further
hadronic decays like $B\to \rho K,\pi K^*,\rho K^*$ are used to impose
additional constraints at the $2\,\sigma$ level. The result of the fit is
used to study the decays $\bar B_s\to\phi\pi^0,\phi\rho^0$ and quantify a
potential enhancement of their branching fractions. Such an analysis,
correlating different hadronic decay modes, is only possible if hadronic
matrix elements are calculated from first principles like in the framework of
QCDF. A method based on flavour symmetries, as it has been used in most studies
of $B\to \pi K$ decays so far, could not achieve this. In particular, the decays
$\bar B_s\to\phi\pi^0,\phi\rho^0$ are not related
to any other decay via $SU(3)_{F}$ so their branching fractions cannot be
predicted in this way.

As far as the $B\to\pi K$ decays are concerned, the new contributions to the EW penguin amplitudes are given by
\begin{eqnarray}
\sum\limits_{i=7,9,7',9'}\rtEWi\, e^{-i\delta_i}&=&(q_7-q_7^{\prime})\,\left[\,(-0.12)^{+0.04}_{-0.05}\,+\,(-0.02)^{
+0.07}_{-0.02}\,i\,\right]\,+\nonumber\\[-0.3cm]
&&(q_9-q_9^{\prime})\,\left[\,0.12^{+0.05}_{-0.04}\,+\,0.02^{+0.02}_{-0.07}\,i\,\right]\,,\nonumber\\[0.3cm]
\sum\limits_{i=7,9,7',9'}\rtEWCi\, e^{-i\delta_i}&=&(q_7-q_7^{\prime})\,\left[\,0.10^{+0.03}_{-0.02}\,+\,0.01^{+0.01}_{
-0.06}\,i\,\right]\,+\nonumber\\[-0.3cm]
&&(q_9-q_9^{\prime})\,\left[\,0.04^{+0.02}_{-0.03}\,+\,(-0.005)^{+0.016}_{
-0.026}\,i\,\right]\,,\nonumber\\[0.3cm]\label{NPRew}
\sum\limits_{i=7,9,7',9'}\rtEWAi\, e^{-i\delta_i}&=&(q_7-q_7^{\prime})\,\left[\,0.03^{+0.04}_{-0.07}\,+\,(-0.06)^{+0.12
}_{-0.01}\,i\,\right]\,+\nonumber\\[-0.3cm]
&&(q_9-q_9^{\prime})\,\left[\,0.007^{+0.003}_{-0.010}\,+\,(-0.006)^{+0.012}_
{-0.003}\,i\,\right]\,.
\end{eqnarray}
From these expressions one draws e.g. the following conclusions:
\begin{itemize}
 \item  parity-symmetric models with $q_7=q_7^{\prime}$ and $q_9=q_9^{\prime}$ do not contribute
 to $B\to \pi K$ at all. Therefore such a scenario cannot
 solve the $\Delta A_{\textrm{CP}}$ discrepancy.
\item The contributions $\tilde{r}_{\textrm{EW},\,7^{(\prime)}}$
 and $\tilde{r}_{\textrm{EW},\,9^{(\prime)}}$
 tend to cancel each other. Hence in the scenarios with
 $q_7=q_9$ and $q_7^{\prime}=q_9^{\prime}$ only a negligible
 new colour-allowed EW penguin contribution is generated.
\item Whereas $\RE(\tilde{r}_{\textrm{EW},\,9^{(\prime)}}^{\textrm{C}})$
 features the typical colour-suppression with respect to
 $\RE(\tilde{r}_{\textrm{EW},\,9^{(\prime)}})$,
 this pattern is not obeyed by the $q_7^{(\prime)}$ terms.
\item The annihilation coefficient $\tilde{r}_{\textrm{EW},\,7^{(\prime)}}^{\textrm{A}}$
 develops a large imaginary part. In scenarios with non-vanishing $q_7^{(\prime)}$
 this term gives the dominant contribution to $\Delta A_{\textrm{CP}}$.
\end{itemize}

From eq.~(\ref{g10}) we see that the $\Delta A_{\textrm{CP}}$ discrepancy can be
solved either through $\rtEW$ or through $\rtEWA$. Except for parity-symmetric
models, any scenario with at least one of $q_{7,9}^{(\prime)}$ different from zero
can achieve such a solution.
The minimal $|q|$\,-\,value needed to reduce the $\Delta A_{\textrm{CP}}$
tension below the $1\,\sigma$ level varies among different scenarios. One finds
e.g. $|q_7|\gtrsim 0.3$ for a model with only $q_7\not=0$, $|q_9|\gtrsim 0.8$ for a model with
only $q_9\not=0$ and $|q_7|=|q_9|\gtrsim 0.4$ for a model with $q_7=q_9$ and the primed coefficients being zero. The
fact that in the $q_7=q_9$ case only a small NP contribution is needed, in
spite of the absence of $\rtEW$, demonstrates the importance of the annihilation
term $\rtEWA$. Finally, we like to stress that the solution of the
$\Delta A_{\textrm{CP}}$ discrepancy via a minimal $|q|$\,-\,value requires
the adjustment of the phase $\phi$ to a certain value. Realistic scenarios
avoiding such a fine-tuning have larger $|q|$\,-\,values,
typically $|q|\sim 1$.

For the $q_7$-only and the $q_9$-only scenarios, the result of the fit to $B\to\pi K$ data and
the consequences for the $B_s$ decays of interest are shown in figs. \mbox{figs.~\ref{figq7q9},\ref{figq7q9P}}
For details about the fit and the observables used we refer to \cite{Hofer:2010ee}.
We find  that the $B\to\pi K$ and related decays set quite strong constraints
on the parameter space, especially in scenarios where $q_9\neq 0$ or
$q_9^{\prime}\neq 0$. This basically rules out the possibility of
having $|q_i|\gtrsim 5$, i.e.\ NP corrections cannot be much larger
than the EW penguins of the SM. The SM point is always
excluded at the $2\,\sigma$ level as a direct consequence of the
$\Delta A_{CP}$ data. According to the sign pattern in eq.~(\ref{NPRew}),
the $B\to \pi K$ fits of the primed and unprimed scenarios in
\mbox{figs.~\ref{figq7q9},\ref{figq7q9P}} are related to each other
through rotation by $180^{\circ}$. The fit works best in
the $q_9^{(\prime)}$ scenario where the best
fit point is given by
\begin{equation}\label{eq:BestFit}
    |\hat{q}_9^{(\prime)}|\,=\,1.9\hspace{2cm}\hat{\varphi}_{9}^{(\prime)}\,=\,-100^{\circ}\,(+180^{\circ}).
\end{equation}
This parameter point yields a full agreement of all the $B\to\pi K$
observables with the experimental mean values.
In the $q_7^{(\prime)}=q_9^{(\prime)}$ case a plateau of $\chi^2=0$ points
arises due to the large theoretical errors. It turns out that the
$B\to \pi K$ observables are not very sensitive to the
$q_7^{(\prime)}$-only scenarios and so the fit does not work well here.

From \mbox{figs.~\ref{figq7q9},\ref{figq7q9P}}
the enhancement $\Br^{\textrm{SM}+\textrm{NP}}/\Br^{\textrm{SM}}$ of the
$B_s$ branching fractions can be read off with respect to the different
constraint- and fit-regions.
In most scenarios an enhancement of an order of magnitude or more
is possible, especially in those involving $q_7^{(\prime)}\neq 0$.
The fact that large parts of the allowed regions do not overlap with
the SM uncertainty regions is encouraging. It means that, if such NP
is realised in nature, it could be possible to probe it easily.

Exceptions are $\bar{B}_s\to\phi\pi$ for $q_7^{(\prime)}=q_9^{(\prime)}$ and
$\bar{B}_s\to\phi_{L}\rho_{L}$ (with the subscript $L$ denoting longitudinally polarised vector-mesons)
for parity-symmetric models, where no enhancement
is expected. Furthermore, effects in the $q_9^{\prime}$ and the
$q_7^{\prime}=q_9^{\prime}$ scenarios are limited by the small allowed region
resulting from the $B\to \pi K$ fit. Largest effects occur as expected in the
scenarios which are least constrained by $B\to \pi K$, i.e. the single
$q_7^{(\prime)}$ and the parity-symmetric models. Especially in these cases
a $\bar{B}_s\to\phi\pi$ measurement would complement $B\to \pi K$ data and,
while the parity-symmetric models lack the motivation via the
$\Delta A_{\textrm{CP}}$ discrepancy, the $q_7^{\prime}$ setting resolves
it with ease. Moreover, we like to stress that $B\to \pi K$ data alone cannot
distinguish among opposite-parity scenarios because such scenarios generate
equal results for the $B\to \pi K$ observables (for $180^{\circ}$-rotated
parameter points). Therefore an analysis of $B\to \pi K$ should for example
be supported by the analysis of a $PV$ decay, suggesting
$\bar B_s\to\phi\pi^0$ as an ideal candidate.

\section{Analysis of specific NP models}

The results of the previous section raise the question: which
concrete models for NP can provide a large new EW penguin amplitude
without being excluded by present data? In this section we provide
a survey of results obtained analysing a set of well-motivated NP
models, referring to \cite{Hofer:2010ee} for more details.
The main difference with respect to the model-independent analysis
is the possibility of adding constraints from other flavour processes
beyond the hadronic $B$ modes, e.g.\ the semileptonic decay
$\bar{B}\to X_se^+e^-$, the radiative decay $\bar B\to X_s \gamma$ and
$B_s$-$\bar{B}_s$ mixing. These processes usually yield tight constraints
on new flavour structures and it has to be investigated if the effects
in $B\to \pi K$ and $\bar B_s\to\phi\rho^0,\phi\pi^0$ survive these
constraints.

\subsection{The modified $Z^0$-penguin scenario}

\begin{figure}[t]
\begin{center}
  \includegraphics[width=0.75\textwidth]{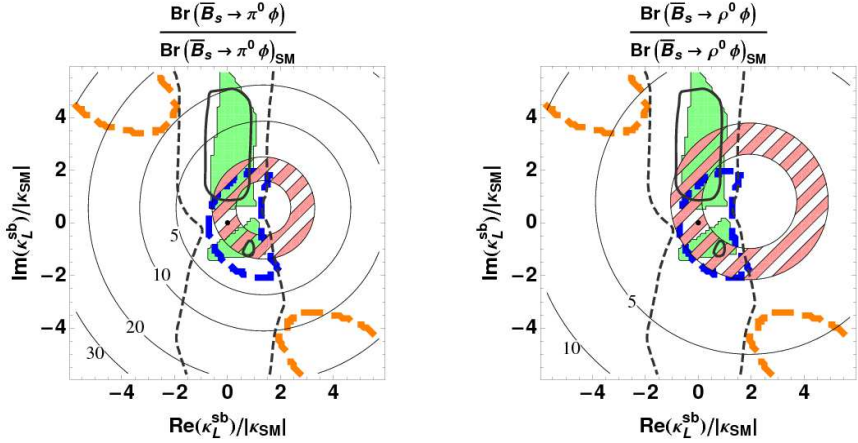}
\end{center}
  \caption{Enhancement factor for the $\bar{B_s}\rightarrow~\phi\rho^0,\phi\pi^0$
  branching ratios with respect to their SM values in the modified-$Z^0$-penguin
  scenario. The green area represents the region allowed by the $2\sigma$
  constraints from all the considered hadronic decays, while the area inside the
  dashed blue line represents the region allowed by the $2\sigma$ constraint
  from semi-leptonic decays. The areas inside the dashed orange line represent the
  parameter values for which the modified-$Z^0$-penguin would explain the
  $B_s$-$\bar{B}_s$ mixing phase}.
  \label{figZ0}
\end{figure}

The simplest class of models with large new contributions to EW penguins
comprises models with a modified $Z\bar{s}b$ coupling. Such
a FCNC coupling can either be generated by integrating out new heavy particles,
e.g.\ in supersymmetric models or fourth-generation models, or it can exist at
tree-level in more exotic scenarios like models with non-sequential quarks.

The model is
described in terms of two parameters, $\kappa_{L,R}^{sb}$, which parameterise
the effective coupling of the $Z^0$ to left- and right-handed quarks, respectively.
A scenario with only $\kappa_L^{sb}\neq0$ scenario shares its most important features
with the $q_9$-only setup of the model-independent study, while the case with only
$\kappa_R^{sb}\neq 0$ is similar to the $q_7$-only case. This expectation
is confirmed by the graphs in fig.~\ref{figZ0}. We only note that we get
a $180^\circ$ rotation due to the signs of $\delta C_9$ and
$C^{\prime}_7$. The main difference is that we now face additional
constraints from semileptonic decays and $B_s$-$\bar{B}_s$ mixing.
The allowed region for the former is given by the interior of the
blue dashed curve, the allowed region for the latter by the orange
areas outside the zone preferred by the $B\to \pi K$ fit. We see that
the $\Delta_s$ anomaly of $B_s$-$\bar{B}_s$ mixing cannot be resolved
in a modified $Z$ scenario when fulfilling at the same time the
semileptonic constraints. However, in models with loop-induced modified
$Z$ couplings the $Z$ exchange contribution to $B_s$-$\bar{B}_s$ mixing
has to be regarded as a subleading effect and therefore the corresponding
constraint does not apply in this case. Focussing on the semileptonic
constraints, we find that they are compatible with the $1\,\sigma$ region
of the $B\to \pi K$ fit for all three cases, but draw the FCNC couplings
$\kappa_{L,R}^{sb}$ to very small values. As a consequence, we expect no
significant effects in $\bar B_s\to\phi\pi^0,\phi\rho^0$.

\subsection{Models with an additional $U(1)'$ gauge symmetry}

\begin{figure}[t]
\begin{center}
 \includegraphics[width=0.75\textwidth]{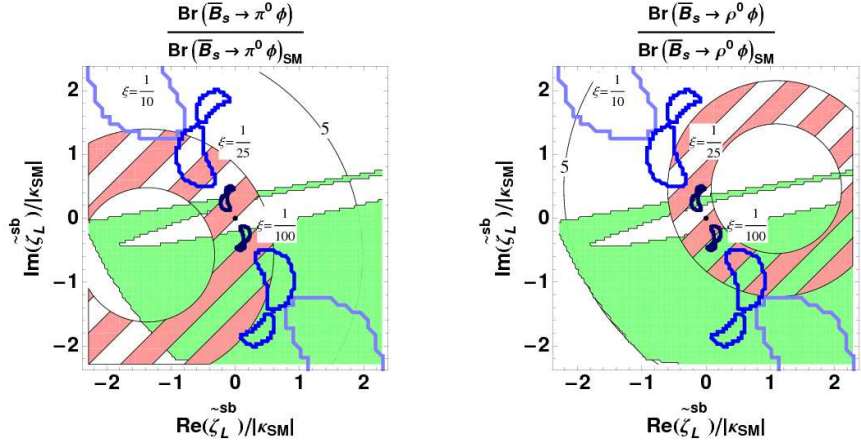}
\end{center}
 \caption{Enhancement factors of $\Br(\bar B_s \rightarrow \phi\pi^0)$
and $\Br(\bar B_s \rightarrow \phi\rho^0)$ for $\tilde \zeta_L^{sb}\neq 0$.
The red-hatched ring corresponds to the SM uncertainty. The green area is
allowed by the $2\sigma$ constraints from all hadronic $B$ decays while the
regions inside the blue lines are compatible with the constraint from
$B_s$-$\bar{B}_s$ mixing. From the biggest to the smallest region they
stand for $\xi=1/10$, $\xi=1/25$ and $\xi=1/100$, respectively.}
\label{figZPL}
\end{figure}

The presence of a heavy $Z^{\prime}$ boson associated with an
additional $U(1)'$ gauge symmetry is a well-motivated extension
of the SM. This additional symmetry has not been invented to solve
a particular problem of the SM, but rather occurs as a byproduct
in many models like e.g.\ Grand Unified Theories, various models
of dynamical symmetry breaking and Little-Higgs models.

The case of interest, in which the flavour
effect is assumed to contribute mainly to the electroweak penguin,
can be described in terms of two parameters, $\tilde\zeta^{sb}_{L,R}$,
corresponding to the properly normalised couplings of the
$Z^\prime$-boson to left- and right-handed quarks, respectively.
The three scenarios with only $\tilde\zeta^{sb}_L\neq 0$, only
$\tilde\zeta^{sb}_R\neq 0$ and with
$\tilde\zeta^{sb}_L=\tilde\zeta^{sb}_R\neq 0$ exactly correspond to
the $q_7$-only, $q_9^{\prime}$-only and the $q_7=q_9^{\prime}$ scenarios
in our model-independent analysis, except for a normalisation factor.
The exclusion regions from the $2\,\sigma$ constraints
and the confidence levels from the fit can be immediately read off
from figs.~\ref{figq7q9} and \ref{figq7q9P}, provided one rescales
the axes by an appropriate normalisation factor and rotates the
pictures by $180^{\circ}$ to take into account a minus sign
in the Wilson coefficients. In this scenario no direct constraints
arise from semileptonic decays, assuming leptophobic $Z^{\prime}$
couplings. The relation to $B_s$-$\bar{B}_s$ mixing moreover
depends on an additional parameter $\xi \,\equiv\,
\frac{g_{U(1)'}^2}{g^2}\,\frac{M_W^2}{M_{Z'}^2}$, which is related
to the overall strength of the $Z^{\prime}$ coupling to fermions,
and to its mass. From the diagrams in fig. \ref{figZPL} we see
that the $B_s$-$\bar{B}_s$ mixing constraint is in general very tight.
It prohibits large effects in $\bar B_s\to\phi\pi^0,\phi\rho^0$
for realistic values of the parameter $\xi\lesssim 1/25$,
which would correspond for example to $g_{U(1)'}\sim g$
and $M_{Z'}\sim 400\gev$.
We find that enhancement of a factor $\sim 5$ is possible in the
$\zeta_L^{sb}$ and $\zeta_R^{sb}$ scenarios whereas no effect can
occur in the $\zeta_L^{sb}=\zeta_R^{sb}$. For $\xi = 1/100$ the
constraints from $B_s$-$\bar{B}_s$ mixing become so strong that
no effect in $\bar B_s\to\phi\pi^0,\phi\rho^0$ would be detectable.
A measurement of a significant enhancement would therefore set a
lower limit on $\xi$, equivalent to an upper limit on the $Z'$ mass.

\subsection{MSSM}

\begin{figure}[t]
 \begin{center}
 \includegraphics[width=0.75\textwidth]{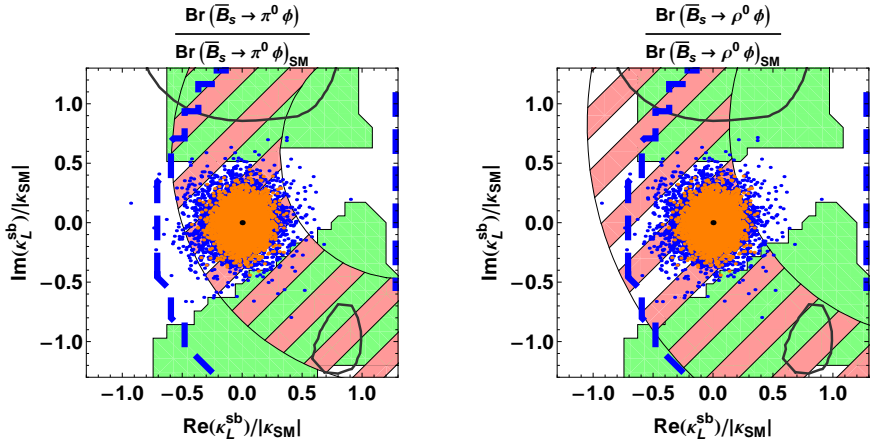}
 \caption{Zoom of the upper plots in fig.~\ref{figZ0}. On top we add
the $\kappa_L^{sb}/|\kappa_\text{SM}|$ values resulting from chargino-induced
flavour-violating Z couplings in a parameter scan.}
\label{enhancementsusy}
\end{center}
\end{figure}

The main contribution to the electroweak penguin
in supersymmetry comes from $Z$ penguins induced by chargino-squark loops.
This scenario is equivalent to a modified $Z^0$ penguin
model with only $\kappa_L^{sb}$ different from zero. Constraints from flavour observables and squark
masses, however, strongly constrain the parameter space available. In
Fig. \ref{enhancementsusy} we provide a zoom of the $\kappa_L^{sb}$
scenario with a superposition of points obtained from a scan of the
MSSM parameter space. The constraints from
squark and chargino masses and from $\Br(\bar B\to X_s\gamma)$
are taken into account. One sees that these points
are not able to decrease the $\Delta A_\text{CP}$  discrepancy
below the $2\,\sigma$ level, nor to create a large enhancement
in the $B_s$ decays branching ratio.

\section{Conclusion}

We have studied the possibility of probing isospin-violating NP
in hadronic $B$ decays. Our analysis is motivated by discrepancies found in
$B\to \pi K$ decays. In particular, the $2.5\,\sigma$ discrepancy found
in the observable $\Delta A_{\rm CP}$ can be interpreted as a sign of NP in
the EW penguin sector of the theory. Since this discrepancy alone is not
able to provide evidence of NP, we have proposed here to consider the
decays $\bar B_s\to\phi\pi^0$ and $\bar B_s\to\phi\rho^0$ as well, which are
extremely sensitive to EW penguins. First, we have demonstrated in a model-independent
analysis that the solution of the $B\to \pi K$ discrepancy is obtained with
an additional NP contribution to the EW penguin operators
$Q_{7}^{(\prime)},...,Q_{10}^{(\prime)}$ of the same order of
magnitude as the leading SM coefficient $C_9^{\textrm{SM}}$. Then,
we have studied the corresponding enhancement of the
$\bar B_s\to\phi\pi^0,\phi\rho^0$ branching ratios for various
scenarios, taking into account constraints from other hadronic
$B$ decays. The results show that in many cases a large enhancement
of about an order of magnitude is possible. Exceptions are
parity-symmetric models, which have no impact
on the $VV$ decay $\bar B_s\to\phi\rho^0$, and scenarios with
(approximately) equal contributions to $C_7^{(\prime)}$ and
$C_9^{(\prime)}$, which cancel in $\bar B_s\to\phi\pi^0$.

The last part of the analysis concerns a survey of concrete NP
models such as modified $Z^0$ penguin, a model with an
additional $U(1)^{\prime}$ gauge symmetry and the MSSM.
In such models additional constraints arise from the semileptonic decays
$\bar{B}\to X_s e^+e^-$ and $\bar{B}\to K^* l^+l^-$ and from
$B_s$-$\bar{B}_s$ mixing. While the solution of the
$\Delta A_{\textrm{CP}}$ discrepancy is in some cases still possible,
like in the modified-$Z^0$-penguin scenario or in a model with an
additional $Z'$ boson, the enhancement of
the $\bar B_s\to\phi\pi^0,\phi\rho^0$ decays is usually strongly
constrained by either the constraint from semileptonic decays or
from $B_s$-$\bar{B}_s$ mixing. Last, we find that the new
contribution to the EW penguins in the MSSM is always marginal
and cannot solve the $\Delta A_{\textrm{CP}}$ discrepancy,
nor create a large enhancement of the $B_s$ decays considered
here.

\subsection*{Acknowledgements}

L.V. would like to thank
the organizers of Discrete 2010 for putting together an exciting
conference in a very pleasant environment.

\end{document}